\documentclass[preprint]{elsarticle}
\journal{International Journal of Forecasting}

\usepackage[utf8]{inputenc}
\usepackage{mathtools, amsfonts, bbm}
\usepackage[british]{babel}
\usepackage{hyperref, booktabs, etoolbox}
\usepackage[super]{nth}

\newcommand{\Z}{\ensuremath{\mathbb{Z}}}
\newcommand{\R}{\ensuremath{\mathbb{R}}}

\newcommand{\E}{\ensuremath{\mathbb{E}}}
\newcommand{\V}{\ensuremath{\mathbb{V}}}
\renewcommand{\Pr}{\ensuremath{\mathbb{P}}}

\newcommand{\ind}[1]    {\ensuremath{\mathbbm{1}_{\{#1\}}}}
\newcommand{\Nb}		{\ensuremath{\mathcal{{NB}}}}

\newcommand{\etal}		{~\emph{et~al.}}
\newcommand{\ts}        {\textsuperscript}

\date{\today}

\begin{document}

\begin{frontmatter}

\title{A white-boxed ISSM approach to estimate uncertainty distributions of Walmart sales}

\author[lok]{Rafael de~Rezende\fnref{vu,tm}}
\ead{derezende.rafael@gmail.com}
\ead[url]{www.linkedin.com/in/rafael-de-rezende}

\author[lok]{Katharina Egert\fnref{db}}
\ead{k@egert.org}
\ead[url]{www.linkedin.com/in/kegert}

\author[lok]{Ignacio Marin\fnref{by}}
\ead{ignacio.marin.eiroa@gmail.com}
\ead[url]{www.linkedin.com/in/ignacio-marin-eiroa}

\author[cnh]{Guilherme Thompson\corref{cor1}\fnref{dt}}
\ead{raposo.thompson@gmail.com}
\ead[url]{www.linkedin.com/in/guilhermethompson}

\cortext[cor1]{Corresponding author}
\fntext[tm]{Team leader}

\address[lok]{Lokad, 83 -- 85 Boulevard Vincent Auriol, 75013 Paris, France (Past)}
\address[cnh]{CNH Industrial, 3 Avenue des Meuniers, 60330 Le Plessis-Belleville, France (Past)}
\address[vu]{Vent--unique.com, 9 -- 11 rue Jacquard, 93310 Le Pré-Saint-Gervais, France}
\address[db]{DB Fernverkehr AG, Edmund-Rumpler-Stra{\ss}e 3, 60549 Frankfurt am Main, Germany}
\address[by]{Blue Yonder, 83-85 Av. de la Grande Arm{\'{e}}e, 75016 Paris, France}
\address[dt]{Dynatrace, 33 Quai Alphonse le Gallo, 92100 Boulogne-Billancourt, France}

\begin{abstract}
We present our solution for the M5 Forecasting - Uncertainty competition. Our solution ranked 6\ts{th} out of 909 submissions across all hierarchical levels and ranked first for prediction at the finest level of granularity (product-store sales, i.e.\ SKUs). The model combines a multi-stage state-space model and Monte Carlo simulations to generate the forecasting scenarios (trajectories). Observed sales are modelled with negative binomial distributions to represent discrete over-dispersed sales. Seasonal factors are hand-crafted and modelled with linear coefficients that are calculated at the store-department level.
\end{abstract}


\end{frontmatter}

\section{Introduction}
The M5 forecasting competition~(\emph{M5}) took place from 2 March to 30 June 2020 on Kaggle~\cite{m5-kaggle}. The challenge was to predict future sales of Walmart products based on past sales. The competition was organised in two parallel challenges. In the first challenge, participants were asked to provide 28 days ahead point forecasts. In the second challenge, a series of quantile estimates for the same period. An overview of the competition and a summary of the conclusions can be found in the \emph{M5}~compendium by Makridakis\etal~\cite{m5-acc-paper, m5-un-paper}.

This article focuses on the latter challenge, for which participants were asked to provide 28-day probabilistic forecasts for the corresponding median and the 50\%, 67\%, 95\%, and 99\% central prediction intervals \cite{m5-rules}. In this challenge, our method ranked 6\ts{th} across all hierarchical levels and first for prediction at the finest level of granularity (product-store sales, level~12). Our approach is conceived mainly to model product-store sales, as this is the most relevant for supply-chain decisions. More details about the competition and benchmarks can be found in the Annex.

We opted for pure statistical models (or rather \emph{structured} models, \cite{barker}) instead of exploring machine-learning frameworks. Despite the recent advances in forecasting using machine learning (i.e.\ ``unstructured'' models by contrast), no approach has yet emerged as an incontestable best practice (see Makridakis\etal~\cite{statsml1} and Green and Armstrong~\cite{Green}, and, more recently, Spiliotis\etal~\cite{statsml2}). In contrast to unstructured methods, the execution time of our approach scales linearly with the amount of time-series as it is almost auto-regressive. Also, our model can be used to produce discrete probability distributions for arbitrarily large horizons.

Our approach is based on both robust statistical modelling and manual tailoring of industry-specific factors. We use an Innovation State Space Model~(ISSM) with exponential smoothing combined with negative binomial distributions for modelling sales at product-store level (level~12). In the supply chain context, using exponential smoothing (damping) means that the long-term demand is expected to fall back to a constant level despite any previous transitory disturbances Chapados~\cite{chapados}. Our model is an adaptation of the multiplicative seasonal Holt and Winters model -- without trend -- and the ETS(A,N,M) model for producing probabilistic sales forecasting via a parametric univariated discrete distribution (see Hyndman~\cite[Sections 2.5 and 16.1]{hyndman}).  However, contrarily to these models, our model is not strictly speaking auto-regressive because our seasonality multipliers are not calculated at the time-series level. 

The linear factors (seasonality multipliers) are used to account for the cyclicities commonly observed in retail, most notably weekly, monthly, and annual seasonalities and the impact of quasi-cyclical events (e.g.\ Super~Bowl or Easter). These multipliers are obtained by analysing the sales patterns at store-department level (level~9) and then pushed down to the product-store level. In our model, these cycles are attenuated by the exponential smoothing of the ``de-seasonalised'' observations. This allows for the trends to be correctly captured by the model while sharp adjustments can be applied using the multipliers. For similar discussions about using factors to work with out-of-stock data see Seeger\etal~\cite{Seeger}, and see Kolassa~\cite{kolassa} for a discussion about day-of-the-week factors. The aggregated sales distributions (and, thus, target quantiles) for product-state level (level~11) and product level (level~10) are obtained by overlaying the trajectories at product-store level (level~12) while assuming they are independent. The sales distributions at the other levels, levels 1 to 9, were obtained by the same method described for level~12, but their amplitude vectors were based on their own historic sales.

Our model is suitable for the vast majority of the concerned series in the competition and for retail in general, especially at product-store level. The negative binomial distribution is an effective way to describe the sales fluctuations inherent to retail (Agrawal and Smith~\cite{agrawal}) and is flexible enough to represent different types of demand (Syntetos\etal~\cite{SyntetosBoylan}). This is mainly due to its capacity to handle over-dispersion common in intermittent sales, which are series characterised by high volume of zero-observations. The negative binomial distribution is repeatedly cited as a suitable choice for modelling this type of sales pattern (see for instance Kolassa~\cite{kolassa}, Chapados~\cite{chapados}, Snyder\etal~\cite{snyder}, Syntetos~\cite{Syntetos2011} and Salinas\etal~\cite{DeepAR}). The negative binomial distribution has only two parameters, mitigating over-fitting while capturing the most important information and keeping the model interpretable. In the context of the \emph{M5}~competition, 73\% of the time-series at product-store level (level~12) were intermittent, while 17\% were lumpy, 7\% were smooth, and 3\% were erratic; these shares are not unique to the \emph{M5} dataset \cite{theodorou2021} but are rather common in retail. The recent results by Ziel~\cite{ziel} confirm the negative binomial model as the best individual model (against Poisson, Geometric, Waring, Double Poisson, Generalized Poisson and zero{-}inflated Poisson) for forecasting at product-store level.

\subsection{Related work}
Sales forecasting intersects several theoretical and practical disciplines from statistics and optimisation to numerical techniques and supply chain. Essentially, our work is inspired by the models of non-stationary count time series with over-dispersed data (variance-to-mean ratio, or index of dispersion, larger than $1$), which have been tested and validated in the supply-chain context (see Hyndman~\cite[Section 16.1]{hyndman}). These models are strongly connected to the study of intermittent sales forecasting. 

In the context of observation-driven time-varying models, the Poisson distribution is the natural candidate because its mean (intensity) parameter can be adjusted to account for time-dependent changes, see for instance Wei{\ss}~\cite{weiss2009} and Fokianos\etal~\cite{Fokianos}. This modelling choice has the highest interpretability and usability because point-forecasts can be ``directly" deployed as the time-varying intensity parameter. However, the Poisson distribution fails to model empirical applications where the variance is larger than the mean - a situation commonly seen in the supply chain context. To circumvent this, we use the negative binomial distribution, a bi-parametric discrete distribution defined by its mean and variance-to-mean ratio which is strictly larger than $1$. Note that the Poisson distribution is a \emph{limiting case} of the negative binomial distribution when the variance-to-mean ratio approaches $1$. 

Our work is inspired by the forecasting approaches in Snyder\etal~\cite{snyder},  Seeger\etal~\cite{Seeger}, Chapados~\cite{chapados}, and Salinas\etal~\cite{DeepAR}. The \emph{Dynamic NegBin Undamped}~(DNBU) model in Snyder\etal~\cite{snyder} is closely related to our structural modelling choice, as it also uses exponential-smoothing updating (dynamic damping) and interprets observations as emissions of a negative binomial distribution. We built upon their model by including linear factors (seasonality multipliers) to account for exogenous variations and seasonalities (see Section~\ref{Sec: spices}). We also add the state vector seed (initialiser) as part of our optimisation program, while Snyder\etal~\cite{snyder} use the mean of the observed sales. In Kolassa~\cite{kolassa} it is argued that the ``poor" performance of the DNBU observed in his experimental setup can be explained by the lack of causal factors. We include such causal factors (seasonality multipliers) in our model. In Svetunkov and Boylan~\cite{SB-ISSM}, the DNBU is outperformed by their intermittent state-space models, which separates the forecast of the demand sizes and the forecast of periods with non-zero demand. This is an adaptation of the Croston's model. However, this performance enhancement is due to the increased complexity of their intermittent state-space models, which requires more parameters (four to six) while the DNBU and our approach are a bi-parametric model.
In Seeger\etal~\cite{Seeger}, the authors combine machine-learning techniques with a state-space model. Their model decomposes the processes governing low-sales emissions including no or only one item sold in a day from the process emitting two or more sales in a day. The latter process is then analysed using a Poisson distribution. The authors use machine learning methods to account for exogenous variables such as date-related properties and out-of-stock events. 
The approach in Chapados~\cite{chapados} is similar, but based on a zero-inflated negative binomial distribution.
In our model, we use a (pure) negative binomial distribution in combination with linear factors to account for seasonalities and important calendar events without using machine learning which makes our model more transparent and interpretable.
The discrete model in Salinas\etal~\cite{DeepAR} is the closest to ours, but they employ a deep-learning structure to estimate the non-linear parameters (with cross-learning) of a negative binomial distribution. Contrarily to Salinas\etal~\cite{DeepAR}, our approach results in the fabrication of a numerically and operationally stable, easy to interpret (white-boxed) forecasting model that is ready to be deployed in production systems. The linear factors in our model can be easily adjusted to account for changes and patterns omitted or opaque in data.

Modelling time-varying count data with a negative binomial distribution is not restricted to sales forecasting. See for instance Davis and Wu~\cite{DavisWu}, Zhu~\cite{zhuNeg}, and Chen\etal~\cite{chen}. 
Examples of other studies that have modelled count data but with different distributions are for instance Zhu~\cite{zhuPoi}, where the Conway–Maxwell–Poisson distribution is used to handle over- and under-dispersion, and Gorgi~\cite{Gorgi}, where a beta–negative binomial distribution is used to increase the model robustness to extreme observations. The score driven beta-negative binomial model in Gorgi~\cite{Gorgi} can be seen as a more flexible version of our model, with one extra parameter to capture the effects of extreme events.
These studies are part of the much broader framework of integer-valued generalized autoregressive conditional heteroscedastic models (INGARCH, see Wei{\ss}~\cite[Chapter 4]{weiss2009} for an overview).
See Cameron and Trivedi~\cite[Chapters 3, 4 and 7]{camerontrivedi} for a comprehensive overview of count data modelling.

\section{The Model} \label{Sec: engine}
In this section, we outline the model used in the \emph{M5} competition. The model adopts a probabilistic 1{-}day ahead perspective. The final many-day ahead estimates are obtained through an iterated generative process where trajectories are created via Monte Carlo sampling.

\subsection{Innovation State Space Model~(ISSM)}
We use a multi-stage state-space model. The state vector is updated using exponential smoothing, and the observations are modelled as `independent' negative binomial distributed observations. This model can be seen as a ETS(A,N,M) model adapted to model discrete distributions via the negative binomial distribution (see Hyndman~\cite{hyndman} and Lipton~\cite{issm}). The underlying hypothesis is that sales are observations (i.e.\ samples or emissions) of `independent' negative binomials distributions whose mean and variance are updated as a function of the previous state. We also use a vector $l = (l_1, \ldots, l_T)$ to encode latent information about the level, seasonality and other factors which impact the sales. This is thoroughly described in Section~\ref{Sec: spices}.

\subsection{A parametric distribution for 1-day ahead sales}
We model the 1-day ahead sales through the negative binomial distribution, see Hilbe~\cite[Chapter~5]{negbin}. The negative binomial is a natural extension of the Poisson distribution to model over-dispersed data, see Cameron and Trivedi~\cite[Chapter 4]{camerontrivedi} or Davis\etal~\cite[Chapter 1]{davis} for a comprehensive overview. We model the negative binomial distribution as a function of its mean and index of \emph{over-dispersion} (variance-to-mean ration plus $1$), similarly to Snyder\etal~\cite{snyder} and Salinas\etal~\cite{DeepAR}.

The random variable $X$ is said to follow the negative binomial distribution $\Nb(\lambda,\theta)$, with $(\lambda, \theta) \in \R_{+}^{2}$, if, for $k \in \Z_{+}$, 
\begin{align*}
    \Pr(X = k \,|\, \lambda,\, \theta) =
        \frac{\Gamma(\lambda/\theta + k)}{\Gamma(\lambda/\theta) \; k!}
         \cdot \left(\frac{1}{\theta+1}\right)^{\lambda/\theta}
         \cdot \left(\frac{\theta}{\theta+1}\right)^{k},
\end{align*}
where $\Gamma(\cdot)$ is the Gamma function \cite{negbin}. The mean of this distribution is $\E[X] = \lambda$ and the variance is $ \V[X] = \lambda(\theta +1)$, thus greater than the mean.

We allow the mean of the negative binomial at a given time $t$, denoted by $\lambda_t$, to change over time to account for structural changes in the demand distribution. Its evolution equation is presented in the next section. We assume that the recorded sales at day $t$, denoted by $y_{t}$, follows
\begin{align*}
    y_{t} \,|\, \lambda_{t} ,\, \theta \,\sim\, \Nb(\lambda_{t} ,\, \theta).
\end{align*}
Note that the variance scales linearly with the mean parameter due to the over-dispersion parameter being fixed.

The optimal parameters for our model are those which maximise the likelihood with respect to the observed (past) values. For the sequence $y = (y_1, \ldots, y_T)$, the likelihood is:
\begin{align*}
    \mathcal{L}(y \,|\, \lambda ,\, \theta) = \prod_{t=1}^{T} \Pr(X = y_t \,|\, \lambda_t,\, \theta),
\end{align*}
where $\lambda = (\lambda_1, \ldots, \lambda_T)$ is the vector of means of the time-varying negative binomial distributions. We use the log-likelihood function instead of the product form to avoid potential computational problems (Cameron and Trivedi~\cite[p.~82]{camerontrivedi}), which can be written as:
\begin{multline}\label{Eq: loglik}
    \ell(y \,|\, \lambda ,\, \theta) = 
        \sum\limits_{t=1}^{T}
      \Phi \left(\frac{y_t + {\lambda_t/\theta}}{(y_t + 1)\lambda_t / \theta} \right) \\
    - \lambda_t / \theta \cdot \log(1+\theta)
    - y_t \cdot \log(1+1/\theta),
\end{multline}
where $\Phi(\cdot)$ is the LogGamma function.

\subsection{Training the model}
Our model requires four parameters: $\alpha$, the smoothing factor controlling the weight given to past observations, $\theta$, the over-dispersion index of the underlying  distribution, $z_T$, the state at the date of the last observation, and the amplitude vector $l = (l_1, \ldots, l_T)$. The three first parameters are learnt through the following process:
\begin{align*}
    && \min_{z_T, \, \alpha ,\, \theta} & \; && -\ell(y \,|\, \lambda ,\, \theta)  \\ 
    && \textrm{s.t.} & \; && z_{t+1} = \alpha \cdot y_t/l_t + (1 - \alpha) \cdot z_t \textrm{ and} \\
    &&&&& \lambda_t = z_{t} \cdot l_t,
    \textrm{ for }  1 \leq t \leq T, \\
    &&&&& z_T ,\, \theta > 0, \textrm{ and } 0 \leq \alpha \leq 1,
\end{align*}
where $\ell(\cdot)$ is the objective function given by Equation~\eqref{Eq: loglik}. Considering the small search space, this optimisation is done via grid search. Note that optimisation is a technical challenge in Seeger\etal~\cite{Seeger}, which we circumvent by the grid search without compromising  the performance of our model. We initialise the sequence $z_{T}, z_{T-1}, \ldots, z_{0}$, using $z_T$ as seed. The amplitude vector is manually constructed to reflect the seasonalities as discussed in Section~\ref{Sec: spices}.

\subsection{A state-space generator for many-day ahead sales}
Once the optimal parameters $\alpha$, $\theta$, and $z_T$ are found, we convert the 1-day ahead probabilistic model of the sales into a many-day ahead forecasting via Monte Carlo sampling (see Hyndman~\cite[Chapters 6, 15 and 16]{hyndman}).
For the forecasting horizon $H$, we simulate the possible trajectories governed by the following system of equations:
\begin{align}\label{Eq: MonteCarlo}
	\begin{cases}
	    z_{t+1} = z_{t} - (z_{t} - y_{t} / l_{t}) \cdot \alpha \\
		y_{t+1} \sim \Nb(z_{t+1} \cdot l_{t+1}, \theta)
	\end{cases}
\end{align}
for $T \leq t < T+H$, where $l = (l_{T+1}, \ldots, l_{T+H})$ is the projected deterministic amplitude vector.

The empirical distribution of the demand on day $t$, denoted by $Y_t$, is given by, for $k \in \Z_{+}$, 
\begin{align*}
    \Pr \left(Y_t = k\right) = \dfrac{1}{U} \sum_{u=1}^{U} \ind{y^u_t = k}
\end{align*}
where, $\ind{\cal{\cdot}}$ is the indicator function, $U$ is the number of simulated trajectories, and $u$, $1 \leq u \leq U$, is an index representing a specific trajectory.
For our submission, we used $10,000$ simulations of trajectories to obtain the empirical probability distributions with the desired precision. Our preliminary experiments showed that above this point the gains were marginal.

\section{Manual parameter estimation} \label{Sec: spices}
In this section, we discuss the manual estimation of the amplitude vector (liner factors or seasonality multipliers) $l = (l_1, \ldots, l_{T+H})$, where $T$ is the length of the past observation time-series and $H$ is the forecasting horizon. These factors were crafted based on our supply chain experience, but no prior knowledge specific to Walmart was involved. 

We use the term \emph{manual} to emphasise the fact that our model specification is the result of empirical experimentation and not a programmatic approach which can be agnostically applied to other contexts such as the \emph{automated parameter learning} presented in Seeger\etal\cite{Seeger}. For instance, the seasonality multipliers were calculated at department-store (level~9) and then pushed down to the products at the same store (level~12). This program can easily be modified to model other hierarchical dependencies but such tailoring needs expert intervention. In the \emph{M5} context, we did not have time to test a large number of model specifications. Further experimentation with different model specifications is a natural extension of the present approach.

We assume that the magnitude of the seasonal fluctuations (amplitude) does scale with the average sales, like in the seasonal multiplicative ETS model (see Hyndman~\cite[Section~2.5]{hyndman}). Allowing for multiplicative effects rather than additive effects is sensible in the context of retail as fluctuations are a function of the average sales level, especially considering daily sales at the product-store level (SKU).

\subsection{Modelling seasonality}
We learnt three distinct multiplicative factors: month-of-year, day-of-week and day-of-month (see for instance Hyndman and Athanasopoulos~\cite[Section~7.4]{fpp3}). Our supply chain experience indicates that month-of-year and day-of-week effects are observed in every supply chain data set (also used in Seeger\etal~\cite{Seeger}). The mixed results obtained in Kolassa~\cite{kolassa} highlight the importance of day-of-week factors and illustrate the challenges in using linear regressors for improving probabilistic forecasting models.

The linear multipliers are deterministic coefficients calculated at the store-department level (level~9). We apply the same coefficients to all products-store series (level~12) belonging to the same department and store. Our coefficients are obtained by leveraging the hierarchical dependencies between product-store and department-store where time related events at the department-store affect demand at the product-store. 

The multipliers are obtained by taking the ratio between average sales for a given department-store at dates of interest (e.g.\ Tuesdays, the 23 day of the month, or the month of May) and the average baseline sales at the department-store level across all dates (the objective being to de-seasonalise the sales). These calculations are somewhat equivalent to the regression of the log transformation of the products-store sales on the set of dummies for month-of-year, day-of-week and day-of-month. A careful implementation is necessary to avoid computational related issues, for instance, in the \emph{M5} context, the coefficients smaller than 0.01 were set to 0.01 to modulate vanishing forecasts.

\subsection{The effects of the Supplemental Nutrition Assistance Program}
While the effects of month-of-year and day-of-week are widely observed in the supply-chain context, the Walmart food sales are, in addition, sensitive to day-of-month effects. This day-of-month seasonality can be explained by the effects of the Supplemental Nutrition Assistance Program (SNAP) on consumer behaviour (Dorfman\etal~\cite{snap}). SNAP is a welfare programme for which around 40 million Americans qualify and consists of cash transfers to needy households in support of their food purchases. This program is often referred to as ``food stamps". The date of the cash transfer differs across US states and influences the timing of consumer's food purchasing decision.
The days for which we included special ``SNAP" dummies to reflect these effect were:
\begin{table}[ht]
    \centering
    \begin{tabular}{ l | l }
    	\hline			
    	State & SNAP Day Impact \\
    	\hline	
    	CA & 1, 2, 3, 4, 5, 6, 7, 8, 9, 10 \\
    	WI & 2, 3, 5, 6, 8, 9, 11, 12, 14, 15  \\
    	TX & 1, 3, 5, 6, 7, 9, 11, 12, 13, 15 \\
    	\hline  
    \end{tabular}
\caption{\label{tab:SNAP} SNAP impact on day-of-the-month sales}
\end{table}

\subsection{Quasi-cyclical events}
The effects of fixed-date events such as Christmas or Halloween on sales are easily captured by the seasonality mentioned above. In addition, we learnt parameters describing the effect of events that are nearly but not exactly cyclical, such as for example Easter or Super Bowl, which happen at different dates every year. If two events take place at the same time, that is, if the calendar specifies more than one event, we picked only the one that has previously had the bigger effect (e.g.\ Easter takes precedence over Orthodox Easter). 

All these factors were combined (multiplied) to generate the amplitude vector, past and future.

\section{Submission}
In the \emph{M5} competition, we used $10,000$ trajectories at the product-store level (level~12) to obtain the target quantiles. We assumed sales for different stores to be statistically independent, so we aggregated sales trajectories from level~12 to create the measurements for product-state level (level~11) and overall product sales (level~10). In other words, we first run the level~12 trajectories, which are then summed/overlaid to create levels 11 and 10. The higher aggregations, levels 1 to 9, were handled like level 12. Providing forecasts of sales at levels 10 to 12 has been the bulk of the work, corresponding to about 99\% of the quantities that needed to be estimated.  

\begin{figure}[ht]
	\centering
	\includegraphics[width=1.0\textwidth]{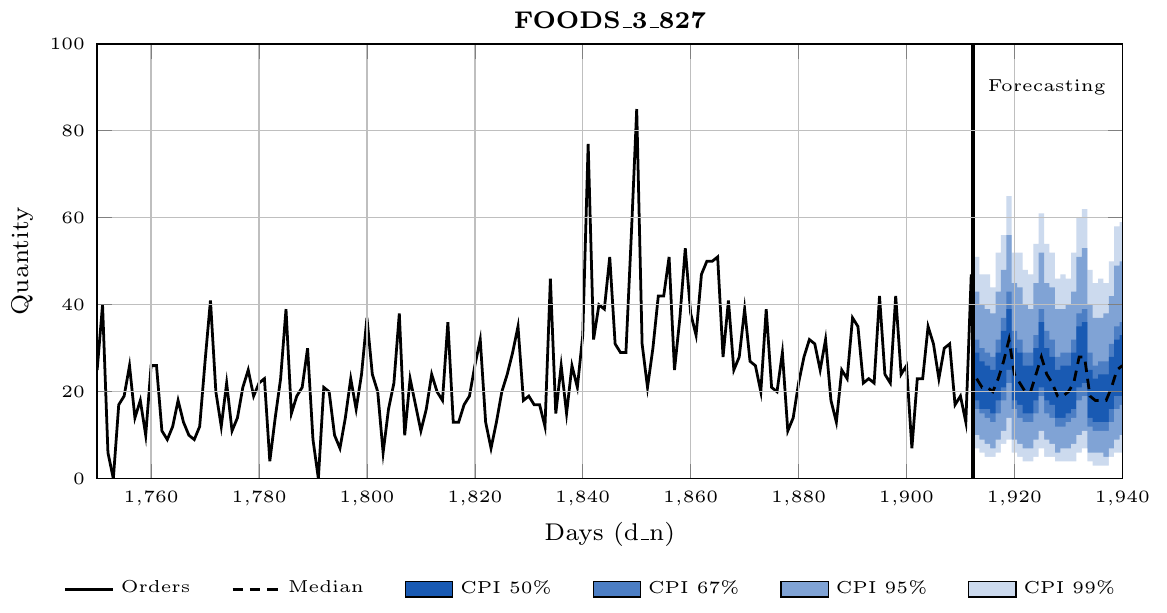}
	\caption{Example of a trajectory at products level (level~10): The different colours depict the central prediction intervals (CPI).}
	\label{Fig: Traj}
\end{figure}

Figure~\ref{Fig: Traj} illustrates the results of our model for product \emph{827} in the \emph{FOODS} category \emph{3} -- already aggregated at level 10. The prediction intervals are depicted using the equivalent quantiles.

\section{Results and discussion}
\subsection{Aftermath of the competition}
Our overall result was a Weighted Scaled Pinball Loss of $0.162$. With this result, we were ranked 6\ts{th} out of 909 participants on \emph{M5} competition. The winner team \emph{Everyday Low SPLices} and runner-up \emph{GoodsForecast} scored $0.154$ and $0.159$ respectively. According to the \emph{M5} survey by Makridakis\etal~\cite{m5-un-paper}, our solution had the best result at the finest level of granularity (level~12 in the competition), commonly referred to as product-store level or SKU level (Stock Keeping Unit). For store replenishment and numerous other problems, the SKU level is the most relevant level. 

Furthermore, our solution had the third-lowest average rank (third best) according to multiple comparisons with the best~(MCB) by Makridakis\etal~\cite{m5-un-paper}. In fact, our method is statistically not significantly different from the solutions by \emph{Kazu} and \emph{slaweks}, which ranked 2\ts{nd} and 4\ts{th} in this analysis.

The results in Makridakis\etal~\cite{m5-un-paper} show that our method outperformed all of the benchmark models (Naive, sNaive, SES, Kernel, ETS and ARIMA) at all levels and all quantiles in the analysed metric. The improvements of our model with respect to the benchmark models are reported in Tables~\ref{tab: bench-level} and \ref{tab: bench-quant} in the Annex.

\subsection{Discussion}
Our solution is based on a state space model with exponential smoothing and negative binomial distributions to model 1-day ahead retail sales. Our model is very simple, reproducible and transparent. The model structure is motivated by our experience with sales forecasting at the product-store level, where the variance is usually larger than the average of the recorded sales. Note that this is the same distribution Amazon's {DeepAR}\footnote{\url{https://docs.aws.amazon.com/sagemaker/latest/dg/deepar.html}} uses for count data, see Salinas\etal~\cite{DeepAR} (and also \cite{snyder, Seeger, chapados}). Our modelling choice is well adapted to forecast sales when variance is larger than the mean (over dispersed data) but it is less adapted to forecast sales at higher aggregation levels (level~1 to 9) where the variance-to-mean ratio is small. Other distributions with more flexibility could have performed better at higher aggregation levels. \emph{Marisaka Mozz}, for example, uses the t{-}student distribution to forecast sales at levels~1 to 9. Another possibility is to inflate or deflate the dispersion parameter as suggested in Seeger\etal~\cite{Seeger}.

Combining the exponential smoothing and the seasonality multipliers allowed us to capture correctly the trends and, yet, adjust quickly our forecasting for sharp transitions related to calendar events exogenous to the time-series. Similar discussions about how to account for exogenous time-varying shocks to sales are present in Seeger\etal~\cite{Seeger} and Kolassa~\cite{kolassa}. We have not accounted for availability (stock-outs) or pricing effects, since we could not properly model such effects with the available data. Including these effects could yield further gains in accuracy. Our approach is tailored to products with a sales history and cannot be used to forecast new SKUs. Using cross-learning, we could potentially extrapolate the parameters for new SKUs from the learnt parameters of similar products in similar locations.

\section{Conclusion}
Our solution is not aligned with the mainstream views of the machine-learning community, which leans heavily on non-parametric and hyper-parametric models. Yet, our solution does not only deliver state-of-the-art accuracy but is also very light both in terms of code complexity and in terms of computing resources. On those latter fronts, we believe to be orders of magnitude leaner than comparatively ranked solutions with the \emph{M5} competition. Furthermore, the white-box design of our approach, where model parameters can be interpreted as capturing the effect of specific events, is of crucial interest for applications in real supply chain environments.

\section*{Acknowledgement}
The authors are grateful to Fotios Petropoulos, Evangelos Spiliotis and the two anonymous reviewers for their constructive suggestions and comments, which helped us to improve the manuscript.

We would also like to thank Julia Mink and Joannes Vermorel for their insightful contributions to this paper.

\bibliographystyle{elsarticle-num}
\bibliography{bibli}

\section*{Annex}

\begin{figure}[ht]
    \centering
    \includegraphics[width=1.0\textwidth]{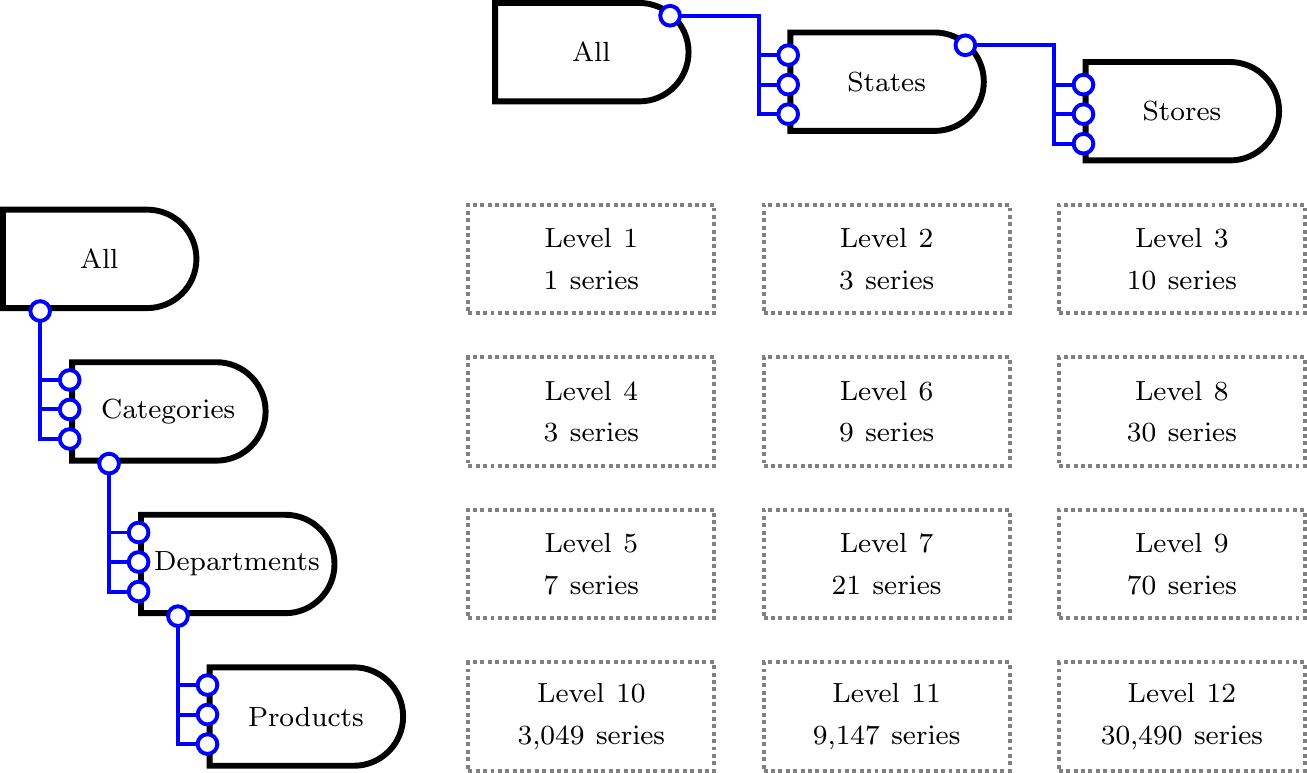}
    \caption{Hierarchical aggregation of the series.}
    \label{fig: logical}
\end{figure}

The data set, provided by Walmart, contains unit sales of various products sold in the USA, organised in grouped time series. It contains unit sales of 3,049 products classified into three product categories (Hobbies, Foods, and Household) and seven product departments in which the above-mentioned categories are disaggregated, see Figure~\ref{fig: logical}. The products are sold across ten stores located in three States, California~(CA), Texas~(TX), and Wisconsin~(WI). The data range from 29 January 2011 to 19 June 2016 (both included). Thus the products have a sales history of 1,941 days.

Thus in total, there are twelve logical aggregation levels for time series, starting from the lowest, unit sales of a specific product in a store -- level 12, to the highest, level 1 -- unit sales of all products aggregated for all stores and states. Level 12, the flattest level, contains 30,490 time series. In Supply Chain data science, this is generally referred to as SKU or product-store level sales data. We produced 42,480 probabilistic time series, characterised by nine quantiles.

The precision of the probabilistic forecasts was evaluated using the \emph{Scaled Pinball Loss~(SPL) function}. The errors per SKU were weighted based on their importance for business, represented by their actual cumulative dollar sales over the most recent 28 observations of the training samples, using the \emph{weighted (SPL) function}, (see \cite{m5-kaggle, m5-rules} for details). Table~\ref{tab: bench-level} and Table~\ref{tab: bench-quant} show the performance by aggregation level and quantile of the our method versus the benchmarks provided by the competition organisers. 

\begin{table}[ht]
	\scriptsize
	\centering
	\begin{tabular}{c|c|c|c|c|c|c|c|c|c|c|c|c}
\toprule
Level & 1 & 2 & 3 & 4 & 5 & 6 & 7 & 8 & 9 & 10 & 11 & 12 \\
\toprule
Our model score & \textbf{.081} & \textbf{.097} & \textbf{.119} & \textbf{.102} & \textbf{.121} & \textbf{.120} & \textbf{.142} & \textbf{.141} & \textbf{.175} & \textbf{.304} & \textbf{.274} & \textbf{.264} \\
\midrule
\emph{Improvement over} &   &   &  &  &  &  &  &  &  &  &  & \\
Naive   & 88\%  & 85\%  & 81\% & 84\% & 81\% & 81\% & 77\% & 76\% & 71\% & 44\% & 45\% & 46\%\\
sNaive  & 56\%  & 52\%  & 45\% & 50\% & 46\% & 46\% & 42\% & 41\% & 33\% & 16\% & 21\% & 27\%\\
SES     & 88\%  & 78\%  & 61\% & 78\% & 77\% & 62\% & 61\% & 55\% & 44\% & 11\% & 11\% & 14\%\\
Kernel  & 84\%  & 79\%  & 75\% & 79\% & 76\% & 73\% & 69\% & 70\% & 62\% & 21\% & 19\% & 15\%\\
ETS     & 42\%  & 39\%  & 30\% & 33\% & 31\% & 33\% & 29\% & 23\% & 15\% & 2\% & 7\% & 12\%\\
ARIMA   & 49\%  & 34\%  & 27\% & 31\% & 28\% & 29\% & 30\% & 21\% & 13\% & 6\% & 8\% &13\%\\
 \toprule
    \end{tabular}
    \caption{Improvement of our method with respect to the benchmark models (in \%) by level.  See Table 2 and Table C1 in Makridakis\etal~\cite{m5-un-paper}.}
    \label{tab: bench-level}
\end{table}

\begin{table}[ht]
	\scriptsize
	\centering
	\begin{tabular}{c|c|c|c|c|c|c|c|c|c}
\toprule
Quantile & 0.005 & 0.025 & 0.165 & 0.25 & 0.5 & 0.75 & 0.835 & 0.975 & 0.995 \\
\toprule
Our model score & \textbf{.020} & \textbf{.056} & \textbf{.201} & \textbf{.256} & \textbf{.332} & \textbf{.283} & \textbf{.229} & \textbf{.061} & \textbf{.017} \\
\midrule
\emph{Improvement over} &   &   &  &  &  &  &  &  &   \\
Naive 	& 43\% 	& 65\% 	& 66\% 	& 61\% 	& 70\% 	& 79\% 	& 80\% 	& 79\% 	& 77\% \\
sNaive 	& 23\% 	& 41\% 	& 46\% 	& 43\% 	& 30\% 	& 30\% 	& 34\% 	& 40\% 	& 37\% \\
SES 	& 31\% 	& 52\% 	& 53\% 	& 49\% 	& 57\% 	& 64\% 	& 64\% 	& 63\% 	& 61\% \\
Kernel 	& 17\% 	& 48\% 	& 63\% 	& 65\% 	& 66\% 	& 65\% 	& 63\% 	& 59\% 	& 43\% \\
ETS 	&  5\% 	& 25\% 	& 29\% 	& 26\% 	& 17\% 	& 17\% 	& 19\% 	& 24\% 	& 29\% \\
ARIMA 	&  5\% 	& 22\% 	& 28\% 	& 27\% 	& 22\% 	& 17\% 	& 14\% 	& 13\% 	& 23\% \\
 \toprule
    \end{tabular}
    \caption{Improvement of our method with respect to the benchmark models (in \%) by quantile. See Table A and Table C2 in Makridakis\etal~\cite{m5-un-paper}.}
    \label{tab: bench-quant}
\end{table}

\end{document}